\UseRawInputEncoding
\documentclass[a4paper,10pt,oneside]{article}
\usepackage[utf8]{inputenc}
\usepackage{icad2023,amsmath,epsfig,times,url,hyperref}
\usepackage[T1]{fontenc}

\title{Design of a multisensory planetarium }

\twoauthors{Stefania Varano} {Istituto Nazionale di Astrofisica, \\Istituto di Radioastronomia\\ Via Piero Gobetti 101 \\ Bologna, Italy   \\ {\tt stefania.varano@inaf.it}}
{Federico Di Giacomo} {Istituto Nazionale di Astrofisica, \\Osservatorio Astronomico di Padova \\ Vicolo dell'Osservatorio 5 \\ Padova, Italy  \\ {\tt federico.digiacomo@inaf.it}}

\begin{document}
\ninept
\maketitle
\begin{sloppy}
\begin{abstract}
We present the design and the prototype of a multisensory planetarium. The goal of this project is to offer a common perceptual and cognitive framework to all users, both sighted, deaf and blind or visually impaired, concerning the experience of observing the night sky, but also to provide to all equal access to scientific data regarding the observed objects, going beyond what can be sensed. \\
The planetarium will consist of a Plexiglas hemisphere on which stars up to the fourth magnitude are represented by a brass bar that, when touched, activates visual, haptic and acoustic stimuli. We mapped the magnitude of stars both on acoustic and on visual stimuli, while the distance of the star from us is mapped on a vibration.\\
All the stimuli have been evaluated in pilot experiments using Plexiglas tablets representing some constellations, to assess their usability, intelligibility and consistency with possible intuitive interpretations. 
\end{abstract}

\section{Introduction}
\label{sec:intro}
Modern astrophysics uses instrumental data for studying objects and phenomena not visible in physical terms, that is they cannot be investigated with the eyes or analogous optical systems. Nevertheless, the fascination of night sky observation still remains one of the main accesses to astronomical contents.\\ 
Since the dawn of time, humans have looked at the night sky and wondered about the nature of what they were seeing. They started arguing possible explanations and eventually seeking for ways to scientifically address their questions, also by investigating the sky using instruments able to observe it beyond what they could sense.\\
The celestial vault is the apparent surface of the sky, which many cultures and myths intend as a sort of physical cover above the observer \cite{plug}. Though being an apparent feature, which existence has been overcome by scientific studies and modern astronomical observations, it still represents a powerful idea, historically, sociologically and culturally grouping humanity. This concept is inaccessible to BVI users, who cannot appreciate the feeling of standing below a black dome checkered with bright points, the stars.\\
Besides the many successful projects using haptic and/or acoustic resources to make astronomy accessible for BVI users, multisensory representations have proven to have a great potential and effectiveness \cite{varano}.\\
Here we present the design, prototyping and preliminary evaluation of a multisensory planetarium, using visual, acoustic and haptic stimuli to convey both the apparent and physical features of stars in the night sky, to offer common perceptual and cognitive frameworks to sighted and BVI users.

\section{DESIGN AND PROTOTYPE}
\label{sec:design}
Stars visible in the night sky have been classified in terms of magnitudes, a feature introduced by Hipparcos in the II century b.C. that is related to the apparent brightness according to the formula ~\ref{eqn:magnitude}
\begin{equation}
  \label{eqn:magnitude}
    m = -2.5 * \log (\frac{F}{F_0})
\end{equation}
where F is the  flux, i.e. the amount of energy per unit time per unit area. Due to the presence of a minus sign and a logarithm, this means that stars of magnitude 1 are ten times  brighter then stars of magnitude 2.\\
The multisensory planetarium will consist of a hemisphere with a diameter of 1.5 meters. Its inner surface will host roughly a hundred brass elements representing the stars up to the fourth magnitude visible in the summer sky. \\
The design process started with the definition of the criteria of selection of a feasible number of stars visible in the night sky and the physical features to be represented in the planetarium. Since the sky changes both according to the geographical location and the moment in which we observe it, we decided to represent the celestial vault visible on June 10, 2023 from Castellaro Lagusello, a small village in the province of Mantua. This choice was prompted by the fact that from 9 to 11 June 2023 there will take place an astronomy festival\footnote{Festival di Astronomia di Castellaro Lagusello: \url{https://www.astronomiacastellaro.eng.oapd.inaf.it/}} in which all the activities will be designed to be multisensory, including not only visual but also tactile and sound elements.\\
Among the approximately 6000 stars visible in the sky, we have decided to put in our multisensory planetarium only the stars up to the fourth magnitude. This means that on the surface of the planetarium users will be able to find about a hundred stars. Through a fisheye projection we will determine the exact position of all sources on the dome which will subsequently be drilled to insert the tactile elements inside. These elements consist in a brass or aluminum bar, of about 5 cm, realized ad hoc for this project (Fig.~\ref{fig:tac}). On top of these elements, we have placed a 3mm LED, while in the terminal part we have inserted a small vibrating motor similar to the one used in smartphones.\\
To test the tactile, visual and acoustic stimuli that we will use in the planetarium, we realized a first prototype. It consists of a Plexiglas tablet showing the Cassiopeia constellation (Fig.~\ref{fig:cass}). The stars of this constellation are represented by five brass elements as described before. To reduce the vibrational effects on the surface of the planetarium which could distort perception, we covered the terminal part of the metal bar with a layer of 3 mm neoprene. We also added a speaker to reproduce the sounds. Finally, we connected all the bars and speakers to an Arduino\textregistered board, the heart of the project. Arduino is an open-source electronics platform based on easy-to-use hardware and software. Arduino boards can read inputs, like touch, and turn it into an output, activating a motor, turning on an LED, playing sounds etc. All inputs are processed by a code, mainly developed in a \textit{C-like} programming language, loaded onto the board.
\begin{figure}[ht]
\centering
\epsfig{figure=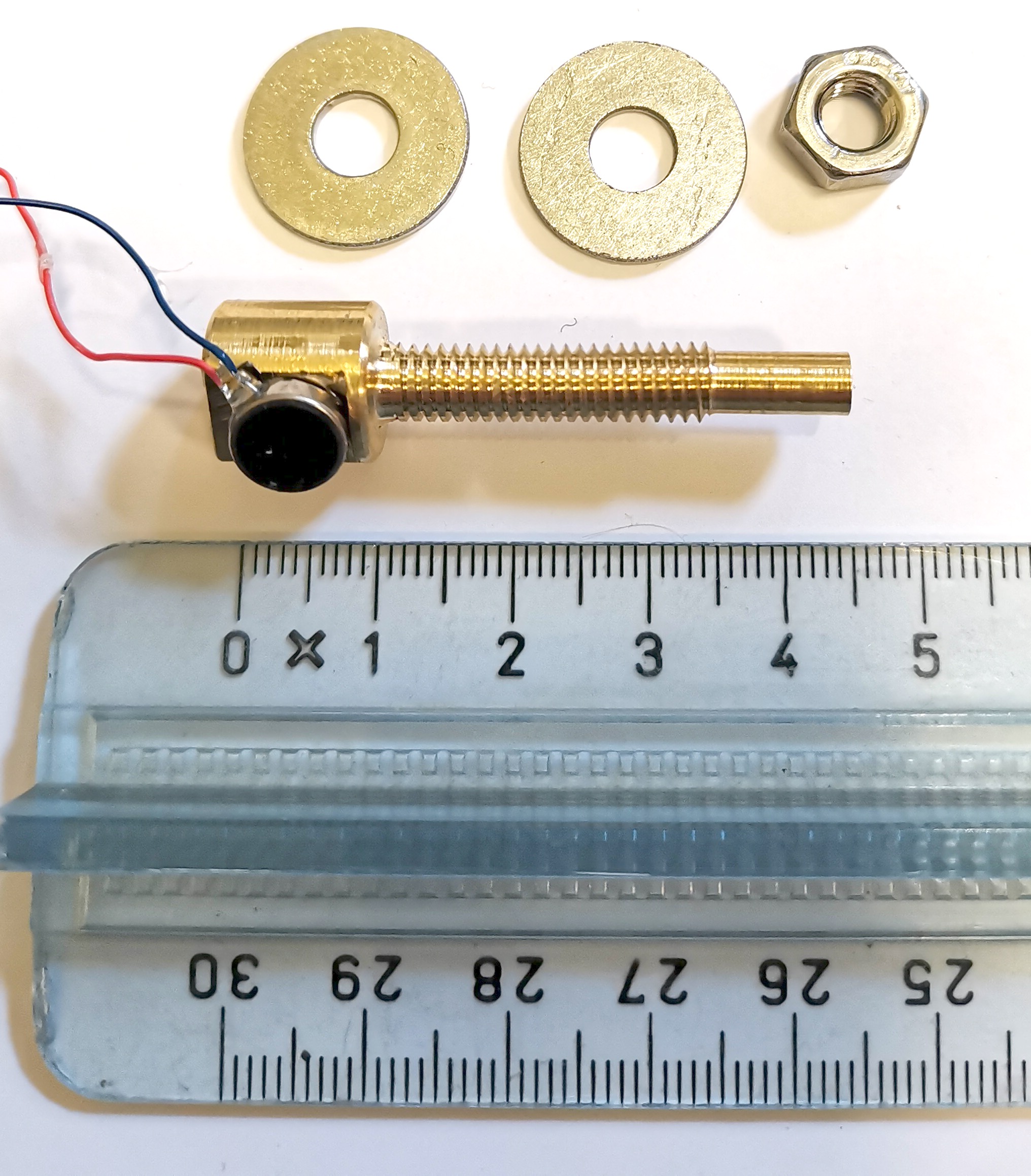,width=0.8\columnwidth}
\caption{Tactile element used to represent the stars both in the prototype and in the planetarium.}
\label{fig:tac}
\end{figure}

\begin{figure}[ht]
\centering
\epsfig{figure=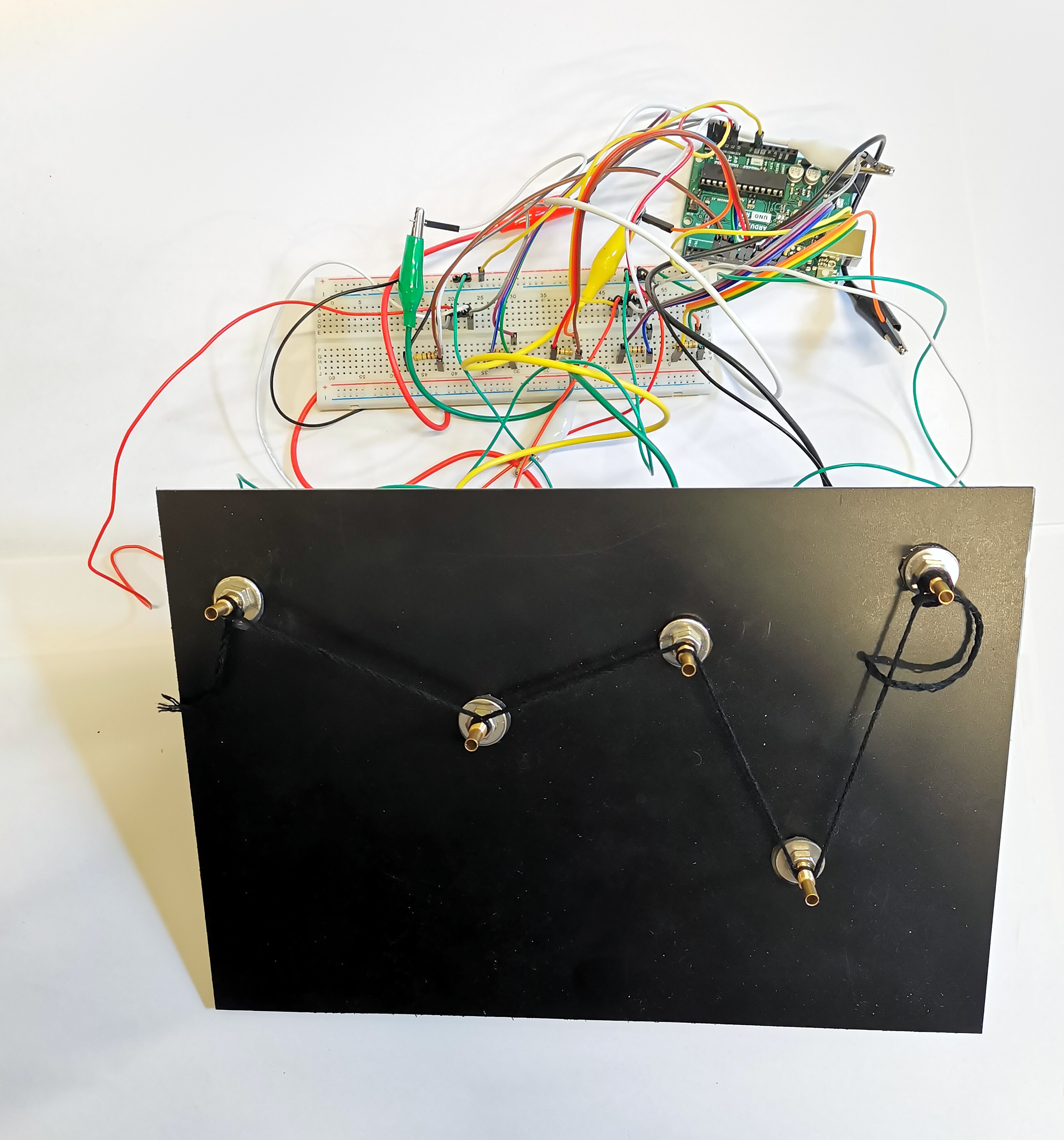,width=0.9\columnwidth}
\caption{In this picture you can see the Plexiglas tablet showing the Cassiopeia constellation. The black thread, joining the brass bars, represents the imaginary line that connects the various stars of the constellation. At the top of the picture you can see the Arduino board with its connection wires.}
\label{fig:cass}
\end{figure}

\subsection{The initial mapping}
\label{ssec:mapping}
The physical characteristics of represented stars mapped onto the planetarium prototype are: the magnitude, position and distance of stars. 
At first, we foresaw a single representation, with the e following mapping:

\begin{enumerate}
    \item  Position: Information about the position of the stars on the celestial vault are provided both as tactile  and visual stimuli, corresponding to the position of the bar on the Plexiglas tablet.
    \item Magnitude: we mapped it both on acoustic and visual stimuli for each source. The visual stimulus was a light from the LED on top of the bar, proportional to the apparent intensity of the represented object.\\
    The acoustic stimulus was a piano note with pitch depending on the apparent intensity of the object. For each value of intensity, we used 4 piano notes at pitches increasing with brightness, as for the main conventions used to map luminosity onto sound for blind users. We chose to use notes one octave apart from each other, to ease the distinction of higher and lower pitches.
    \item Distance: The distance of the star from us was mapped on a vibration, faster/slower for smaller/greater distances. We used a vibrating motor, such as the ones used in smartphones, with a fixed frequency of 240 Hz. In particular we have chosen to increase the pulse interval by 100 milliseconds every 50 light years \footnote{One light year is 9.46 × 10$^{15}$ m.}\cite{wersényi}. In table ~\ref{tab:vibration} we report the stars of the Cassiopeia constellation with the corresponding distance and the frequency of the vibration.\\ 
\end{enumerate}

\begin{table}[h]
    \centering
        \begin{tabular}{lccc}
            \hline
            \noalign{\smallskip}
            Stars & Distance [ly] & Pulse interval [ms] \\
            \noalign{\smallskip}
            \hline
            \noalign{\smallskip}
            $\alpha Cas$  & 228 & 400 \\
            $\beta Cas$ & 54 & 100 \\
            $\gamma Cas$  & 613 & 1000 \\
            $\delta Cas$ & 99 & 200 & \\
            $\epsilon Cas$  & 458 & 800 \\
            \noalign{\smallskip}
            \hline
         \end{tabular}
    \caption{Characteristics of vibrational stimuli for the stars of the Cassiopeia constellation.}
    \label{tab:vibration}
   \end{table}

\section{PILOT TESTS AND FIRST EVALUATION}
\label{sec:pilot}

We conducted a preliminary evaluation session with 2 BVI users aged between 50 and 60, of the \textit{Institute for Blinds “Cavazza”} in Bologna, for assessing the usability of the representation and the understandability of the sensory stimuli.\\
The main aim of this test was to address the validity of the vibrating and acoustic stimuli, and the understandability of their correspondence with the represented physical features, to exclude possible inconsistencies with intuitive interpretations. In this test we used the prototype described in the previous section.\\
In the first step we assessed only the vibration. At first, we asked the users to autonomously explore the tablet and refer aloud their first impressions. They started exploring the entire plan of the tablet, looking for its boundaries and for the objects. Then, they argued for possible geometric symmetries in the representation. Not finding any, they started concentrating on the haptic stimuli, and comparing the vibrations of each object with the others. They correctly argued that two objects were vibrating at the same rhythm, but promptly asked about the meaning of the different vibrations.
Once informed about the fact that the objects they were touching represented stars visible in the night sky and the different vibrations meant different distances of such objects from Earth, they correctly associated faster vibrations with closer objects and slower ones with objects further away.\\
In the second step we also introduced the sounds to describe the different magnitudes. As described in the previous section we use four different octaves in which a higher pitch indicates an apparent brighter star.

\section{PRELIMINARY OUTCOMES AND CONCLUSION}
\label{sec:conclusion}

The feedback on our pilot test was very positive.\\
The tablets proved to be fairly usable and easy to explore.  \\
The presence of a perceptual equivalent of what is inaccessible and perhaps even unimaginable for a BVI user was very much appreciated.\\
The information about the distance was appreciated in a deeper cognitive framework, as extra information based upon scientific data, inaccessible through a simple and naked-eye (or ear) observation of the sky. A different cognitive framework, indeed. \\
This led us to some considerations about the chosen representation.\\
In particular, we realized that the use of sounds one octave apart from each other as substitute stimuli of the different intensities of stars visible in a naked-eye observation was perhaps inappropriate, since the difference between the two representing sounds is much easier to argue than the difference in visual intensity of the stars seen in the sky or even of the light of LEDs representing them. We then decided to map the intensity of light into sounds at different volumes, to offer a more appropriate perceptual equivalent.\\
On the other hand, the presence of stimuli representing the distance, which cannot be perceived with the senses and must be retrieved from astronomical instrumental observations, brought us to mind the opportunity of creating an additional representation, on a different cognitive plane.\\
Such representation, being more quantitative and objective, could offer the users the opportunity to argue the actual represented values. In this representation, we used light and sounds with different duration (1 to 4 secs) to map the four values of represented intensity.\\
The vibration mapping the distance is only included in the second representation. 
This choice allows us to offer an equal perceptual fruition of what can be observed in the night sky without using instruments and telescopes, and on the other hand, a likewise equal perceptual representation of astronomical data, on two different cognitive levels, both accessible to all.

\section{FUTURE WORK }
\label{sec:future}
The first multisensory planetarium was presented at the astronomy festival of Castellaro Lagusello (9 - 11 June 2023).\\ 
In this first version we used the dome in tactile format providing information on the position and apparent magnitude of the stars as described in section ~\ref{sec:design}. Alongside this we will also use some tablets representing the constellation of Cassiopeia, the big and the small dipper and other summer constellations.\\
After evaluating and discussing the usability of this setup and the inserted stimuli (both in the dome and in the tablets), our intention is to insert all the information present in the tablets directly into the dome. In this way it will be possible, by means of a suitable switch, to switch from a purely perceptual representation to a more quantitative one, providing information on apparent distances and brightness.

\section{ACKNOWLEDGMENT}
\label{sec:ack}
We thank Fabio Fornasari, Paola Gamberini and Fernando Torrente for their support and assistance during the pilot test phase and for their comments which helped us significantly improve this project.



\end{sloppy}
\end{document}